\documentstyle[prd,tighten,aps]{revtex}	
\begin{document}
\draft
\twocolumn[\hsize\textwidth\columnwidth\hsize\csname
@twocolumnfalse\endcsname  

\title{  A solution to the zero-hamiltonian problem in 2-D gravity}
\author{  C. P. Constantinidis $^*$ F. P. Devecchi $^{**}$ and
D. F. Marchioro $^{**}$}
\address{$^*$
 Depto. de F\'{\i}sica, 
Universidade Federal do Esp\'{\i}rito Santo, av. F.
 Ferrari-Goiabeiras,\\  cep 29060.900, Vit\'oria-ES, Brazil.}
\address{$^{**}$ Depto. de F\'{\i}sica, Universidade Federal do Paran\'a, 
c.p. 19091, cep 81531.990, Curitiba-PR, Brazil.}
\maketitle

\begin{abstract}
 The zero-hamiltonian problem, present in reparametrization  invariant systems, is
solved for  the 2-D induced  gravity model.  Working with
 methods developed by Henneaux et al. we find sistematically the
 reduced phase-space physics, generated by an {\it effective}
 hamiltonian obtained
 after complete gauge fixing.
\end{abstract}


\vskip2pc]

\section {introduction} 

In  theories of gravitation the canonical hamiltonian is a linear
 combination of constraints,  meaning that after gauge fixing
it turns into a strongly zero quantity. This is in brief the 
 ``zero-hamiltonian problem" (ZH problem), present, for a more
 general  context, in diffeomorphism ({\it Diff}) invariant
models.  

 Two-dimensional (2-D) gravity models has been under intensive investigation
 during the last
two decades \cite{Stro}. The old problem of quantum gravity, black holes
 physics
and string dynamics were tested in
these formulations. In particular, the so called {\it induced } gravity
 model
  was proposed in the eighties
 by Polyakov
\cite {Pol}; Emerging as a  field theory of gravitation after
 integration of the matter fields. In spite of its peculiar features
and succesful quantum formulation, the ZH problem is also present
 in Polyakov's model.

The ZH problem was analysed by Henneaux, Teitelboim and Vergara
 \cite {Hen}
in a broad context. The idea was to construct an extension on the original
 action that is invariant under gauge transformations not vanishing at the
end-points;
   the boundary conditions were therefore modified through the gauge
generators.
The extension mentioned above is related to the physical ({\it effective})
   hamiltonian of the
   theory that
it going to rule the dynamics of the physical degrees of freedom.
 An alternative approach was proposed by Fulop, Gitman and Tyutin 
  \cite {Git}; The main point here is
   that one works in the reduced phase-space. Once
    determined the simplectic structure, after
     {\it complete} gauge fixing, a time-dependent
      canonical transformation is performed, obtaining
       the dynamics generator
 for the physical variables.

In this work we apply these methods to the 2-D induced gravity model
 obtaining the
  dynamics and the {\it effective} hamiltonian in a
  sistematic way, working in reduced phase-space.

 The manuscript is structured as follows. In Section II we make a
  panoramic description of the methods mentioned above to analyse the
   ZH problem.
  In section III we  apply these techniques in  the relativistic
particle model, working in the proper-time gauge. In section IV we discuss
our main interest, the 2-D  induced gravity case. Finally,
 in Section V we display
     our conclusions.

\section { Solving the zero-hamiltonian problem } 

 In this section  we make a 
short review of the techniques
developed in \cite{Hen} (and  partially in \cite{Git}). These procedures
 solve the ZH problem, present 
in 
  ({\it Diff}) invariant  theories.

Given a gauge system with {\it  Diff} invariance we can 
write its action as
(we use finite-degrees-of-freedom's notation for simplicity)

\begin{equation}
S=\int L d\tau = \int ^{\tau _2}_{\tau _1} (P_i \dot Q^i -H_0
 -\lambda _a G^a )d\tau \,\,\, ,
\label{1}
\end{equation}

\noindent where the $P_i$'s ($i=1,..,N$) represent the canonical
 momenta conjugated
 to the coordinates $Q^i$, $H_0$ is the canonical Hamiltonian 
 and the $\lambda _a$ are  Lagrange multipliers. As a consequence of 
the {\it Diff} invariance of the action, the system has a set
 of $a$ ($a=1,..,R)$ first-class 
constraints. They satisfy, by definition,  the Poisson algebra \cite{Dir}

\begin{equation}
\{ G_a,G_b\} =C _{ab}^c G_c\approx 0\,\,\, ,
\label{2}
\end{equation}

\noindent that in the case of gravity theories turns into the
 well-known {\it Diff} algebra \cite{Abda}. There are not second-class
   constraints.

The  presence of the ZH problem is a consequence of the
properties mentioned above. In fact, for {\it Diff} invariant systems the 
total hamiltonian ($H_T$) is a linear combination of  constraints 

\begin{equation}
H_T=H_0+\lambda ^a G_a\approx 0\,\,\, , \label{3}
\end{equation}

\noindent meaning that it is a {\it strongly zero} quantitiy after the
(complete) gauge fixing
procedure. This fact leaves no generator of dynamics
 in the reduced phase-space \cite{Dir} \cite{Hen}.

To solve the ZH problem, Henneaux Teitelboim and Vergara \cite{Hen}
 (and in a different context  Gitman and Tiutyn \cite{Git}
 ) proposed to perform
an extension on $S$ that takes into account  end-point contribuitions.
The action for the paths obeying these open boundary conditions
 ( the gauge
 parameters $\epsilon ^a$ do not vanish at the end points) is

\begin{equation}
S_E= \int _{\tau _1}^{\tau _2}( pq-H_0-\lambda ^a G_a )
 d\tau -[P_i\frac{\partial G}{\partial P_i}-G]^{\tau _2}_{\tau _1}\,\,\, ,
\label{4}
\end{equation}

\noindent with $G\equiv \epsilon ^aG_a$, where the
 $\epsilon $ are the gauge transformation parameters in
  the {\it extended} hamiltonian formalism \cite{Dir} \cite{Deve}
\cite{Deve2}.  
 The new boundary conditions
are given by 

\begin{equation}
\bar Q(\tau _i )\equiv [Qî-\{Qî, G\}](\tau  _i)\,\,\,\,\, i=1,2\,\, ,
\label{5}
\end{equation}

\noindent with  analogous expression for the new momenta $\bar P$.

The correspondent  {\it generating function} ($M$) 
 is related
to the gauge ({\it Diff }) generator. We have

\begin{equation}
M=P_i\frac{\partial G}{\partial P_i}-G\,\,\, ,
\label{6}
\end{equation}

\vskip.5cm

\begin{equation}
S' =S+\int ^{\tau _2}_{\tau _1} \frac{dM}{d\tau} d\tau
\label{7}\,\,\, ,
\end{equation}

\noindent
and the action (\ref{4}) is invariant under  gauge transformations with the
open boundary conditions (\ref{5}). 

\vskip .5cm
It is also possible to obtain a  non-zero {\it effective} hamiltonian function
($\bar H$) in reduced phase-space. After gauge fixing
a new canonical transformation is performed, whose generator ($F$)
is  determined by the form
of the gauge fixing constraints \cite{Git}.
 We can write

\begin{equation}
\bar H= \left[ H_T + \frac {\partial F}{\partial \tau } \right]
 \vert  _{fixed} \,\,\, .
\label{8}
\end{equation}

 \noindent We obtain the last equality  after (complete) gauge fixing,
 meaning that $\bar H$ is the
hamiltonian function for the new variables, in reduced phase-space.
 The equations
 of motion will be

\begin{equation}
\dot {\bar Q} =\{ \bar Q, \bar H\} _D\,\,\,\,\, \dot {\bar P}=
\{ \bar P, \bar H \} _D\,\,\, ,
\label{9}
\end{equation}

\noindent where D denotes the Dirac bracket \cite {Dir} operation. The 
 constraints surface in the new variables 
is  given by

\begin {equation} 
\bar G^a(\bar Q, \bar P) \equiv G^a \left[ P(\bar Q , \bar P), Q(\bar Q, \bar P)
\right]
\label{10} 
\end{equation}                                    

with analogous expressions for the gauge fixing constraints 
(that will turn the weak equalities into strong ones). 

\vskip .5cm

\section {The relativistic particle example}

In this section we take the  instructive relativistic particle
example to apply the methods described in
 the precedent section. Essentially this analysis   was done
  in \cite{Hen}
and partially in \cite{Git}. Here we  reproduce some calculations
 for pedagogical reasons and also
  find correspondent results when  the proper-time gauge
fixing is considered.
\vskip .5cm  

 As  usual, the action is   proportional
 to the world-line length
\begin{equation}
S=-m\int ds \,\,\, ,
\label{11}
\end{equation}

\noindent from which  follows 
 the parametrized ($ \tau $) lagrangian

\begin{equation}
L=-m(-U^{\nu} U_{\nu })^{\frac{1}{2}}\,\,\, ,
\label{12}
\end{equation}

\noindent 
 where the $U's $ are
  the four-velocities ($U^{\nu}\equiv \frac {dx^{\nu}}
{d\tau } $). The metric convention is diag [-1, 1, 1, 1].
As is well known, the action (\ref{11}) is {\it Diff} invariant
($\tau \rightarrow \tilde \tau $).
\vskip .5cm

The total hamiltonian  $H_T$ is easily
 constructed, being proportional to constraints as expected.
  The canonical
  hamiltonian $H_c$ is strongly zero

\begin{equation}
H_T=H_c  + \theta _1 \xi _1   =\theta _1\xi _1 
\,\,\, . 
\label{13}
\end{equation}

\noindent  The model has only one constraint, which
 is evidently
first class ($\theta _1 = p^{\mu } p_{\mu } + m^2 $). $\xi _1
$ is an arbitrary local multiplier. 

The  Hamilton-Jacobi equations of motion read

\begin{mathletters}
\label{14}
\begin{eqnarray}
\frac{dx^{\mu}}{d\tau } & \approx & \{ x^{\mu} ,H_T\} = 2\xi _1  p^{\mu }
\label{mlett:1} \\
 \frac{dp_{\mu}}{d\tau } & \approx & \{ p_{\mu} ,H_T\} = 0
\label{mlett:2}\,\,\, .
\end{eqnarray}
\end{mathletters}

The gauge transformations generated by the first-class constraint
$\theta _1$
are analogous to the equations of motion above since  the model
is ``pure gauge"

\begin{mathletters}
\label{15}
\begin{eqnarray}
\delta _{\epsilon} x^{\mu} &=& \{ x ^{\mu } , G\} =
\{ x^{\mu}, \epsilon \theta _1 \} = 2 \epsilon p^{\mu }
\label{mlett:1} \\
\delta _{\epsilon} p_{\mu } &=& \{ p_{\mu }, G \} = 0 \,\,\, .
  \end{eqnarray}
\end{mathletters}

\vskip .3cm

The {\it generating function } ($M$) is in this case, following
the definition (\ref{6})

\begin{equation}
\lbrace p_{\mu} \frac{\partial G}{\partial p_{\mu}}-G\rbrace ^{\tau _2}_{
\tau _1}= \epsilon (p^2 - m^2)\vert ^{\tau _2}_{\tau _1}\,\,\, . 
\label{16}
\end{equation}

\noindent So, the improved action reads

\begin{equation}
\bar S = S+ \frac{\delta x^0}{2p^0} (p^2 - m^2)
\vert ^{\tau _2}_{\tau _1} \,\,\, ,
\label{17}
\end{equation}

with the new boundary conditions given by

\begin{equation}
X^{\mu} (\tau _i) = [x^{\mu } - \frac{\delta x^0}{p^0} p^{\mu }] (\tau _i)
\,\,\,\, i=1,2 \,\, .\label{18}
\end{equation}

\vskip .5cm

The next step is the reduced phase-space analysis.
The proper-time gauge fixing  condition is given by 

\begin{equation}
\theta _2 = x^0 - \frac{p^0}{m} \tau \,\,\,\, .\label{19}
\end{equation}

Following the standard procedure \cite{Dir}, the Dirac
 brackets for the physical degrees of
 freedom (the ``spatial'' sector $[x_i$ , $p^i]$) are 

\begin{equation}
\{x_i , x_j\} _D  = 0 =
\{ p^i , p^j\} _D \,\,\,\, 
\{ x_i , p^j \} _D = \delta ^j_i \,\, . \label{20}
\end{equation}

\noindent The gauge fixing condition (\ref{19}) gives the form of
 the canonical
 transformation
needed

\begin{mathletters}
\label{21}
\begin{eqnarray}
P_{\mu} &=& p_{\mu}\\
X_i &=& x_i \\
X_0 &=& x_0 - \frac{p_0}{m} \tau
\end{eqnarray}
\end{mathletters}

\noindent As a consequence the constraints in the new
 variables become

\begin{equation}
\bar \theta _1 = \theta _1 \,\,\,\,\,\,\, \bar \theta _2\equiv 0 \,\,\, .
\label{22}
\end{equation}

\noindent

The generator of the canonical transformation is a function of old momenta
, new positions and time (Type $F_3$, see \cite{Gold} )

\begin{equation}
F_3= -X^{\mu } P_{\mu }+ \frac {(P^0)^2}{2m}\tau \,\,\, .
\label{23}
\end{equation}

\noindent So the hamiltonian in reduced phase-space is

\begin{equation}
\bar H= H + \frac{\partial F_3}{\partial \tau} = \frac{(P^0)^2}{2m}\,\,\, ,
\label{24}
\end{equation}

\noindent giving the right Hamilton-Jacobi equations for the proper-time
description

\begin{mathletters}
\label{25}
\begin{eqnarray}
\frac{dx^{i}}{d\tau } = \{ x^{i} ,\bar H \} _D = \frac { p^{i}}{m}
\label{mlett:1} \\
 \frac{dp^{i}}{d\tau } = \{ p^{i} ,\bar H \} _D = 0
\label{mlett:2}\,\,\, .
\end{eqnarray}
\end{mathletters}

\vskip .3cm

\section {The induced 2-D gravity model}
\vskip .5cm

In this part we apply the techniques described in
section II to a field theory; of our interest is  the 2-D induced gravity
 model
 proposed by Polyakov
in \cite{Pol}.

We first want to  find the generator of the local gauge
 transformations ({\it
Diff }). The form variations in the fields are 

\begin{mathletters}
\label{50}
\begin{eqnarray}
\delta \varphi &=& \epsilon ^{\mu } \partial _{\mu } \varphi \\
\delta g^{ \mu \nu } &=& \partial _{\sigma }
g^{\mu \nu } \epsilon ^{\sigma } - g^{\mu \sigma }
\partial _{\sigma }\epsilon ^{\nu }
+g^{\nu \sigma } \partial _{\sigma } \epsilon ^{\mu }\,\,\, .
\end{eqnarray}
\end{mathletters}

\noindent
It is possible  to write the corresponding action as a local functional,
 introducing the
auxiliary  scalar field $\varphi (x)$ \cite{Abda}

\begin{equation}
S=\int d^2x \sqrt{-g} \left(-\varphi \Box \varphi - 
\alpha R\varphi +\alpha ^2\beta
\right) \,\,\, ,\label{51}
\end{equation}

\noindent where $R$ is the 2-D scalar curvature. $\alpha $ and
$\beta $ are  scalars related 
 to a central charge  
(gravity
coupled to matter, before integration) and to a 
 cosmological constant, respectively (
 for details see \cite {Pol}).

The generator of  the local invariances (\ref{50})  must be a linear
 combination
 of the
 first class constraints that arise from (\ref{51})
 (there are not second class constraints). We have

\begin{mathletters}
\label{52}                 
\begin{eqnarray}
\omega _1 = \pi ^{00}\approx &0&  \label{mlett:1}  \\
\omega _2 = \pi ^{01}\approx &0&  \label{mlett:2}\,\,\, .
\end{eqnarray}
\end{mathletters}

\noindent where $\pi^{\mu \nu}$ are the momenta conjugated to the metric
components $g_{\mu \nu}$.
The time-consistency condition \cite {Dir} applied
 to the primary constraints (\ref{52})
 gives two secondary constraints
 
\begin{mathletters}
\label{53}
\begin{eqnarray}
  \phi _1 &=& \frac{1}{2} \left( \varphi ^{'2}-\frac{4}{\alpha
^2}(g_{11}
\pi ^{11})^2-
\frac{4}{\alpha }(g_{11}\pi ^{11})\pi _{\varphi}
 \right. \nonumber \\ 
&-& \left. \alpha 
\frac {g_{11}'}{
g_{11}}+2\alpha \varphi '' + \alpha ^2\beta g_{11}\right) 
\label{mlett:1}\\
\phi_2  &=& \pi _{\varphi } 
\varphi ' -2g_{11}\pi ^{11'}-\pi ^{11}g_{11'}\,\,
 \label{mlett:2},       
\end{eqnarray}
\end{mathletters}

\noindent and no more constraint generations appear. In accord with the
discussion
of section II, the {\it Diff}
invariance implies into a  hamiltonian functional  that is proportional to
$\phi _1$ and $\phi _2$

\begin{equation}
H_c=-\frac{\sqrt {-g}}{g_{11}} \phi _1+ \frac{g_{01}}{g_{11}}           
 \phi _2\,\, .\label{54}
\end{equation} 

Following the Anderson-Bergmann algorithm \cite{Ber}
, the generators of the  gauge  transformations
(\ref{50}) must obey (taking for example the variations of
 the $\varphi$ scalar
 field)

\begin {eqnarray}
 \delta \varphi (x) &=& \{ \varphi (x), G \} \nonumber\\
 &=& \int dy \{ \varphi (x), 
\left( \dot \epsilon ^1 \bar G ^0_1+ \dot 
\epsilon ^0 \bar G ^0_0 + \epsilon ^1 \bar {
 \bar {G_1}} +\epsilon ^0 \bar {\bar {G_0}} \right) \} \nonumber \\
&=& \epsilon ^0 \partial _0 \varphi (x) +\epsilon ^1 \partial _1
 \varphi (x) \nonumber \\  &=& \epsilon ^0 \left( \frac {2\sqrt {-g}}{\alpha }
 \pi ^{11} +
\frac {g_{01}}{g_{11}} \varphi '\right)(x)+ \epsilon ^1
 \partial _1 \phi (x)\,\,\, ,
\label{55}
 \end{eqnarray}

 \noindent where we have used the definition of momentum $\pi ^{11}$

 \begin{equation}
 \label{56}
\pi ^{11 } = \frac{\alpha }{2\sqrt {-g}}\left[ \dot \varphi -
\frac{g_{01}}{g_{11}}\varphi '\right]
 \,\,\, ,
 \end{equation}

  \noindent to put the
time derivatives  in hamiltonian form. The $\bar G$ variables
 are linear
 combinations
of contraints (\ref{53}) and (\ref{52}). To determine their form we see that
$ \bar  G ^0_1$ and $\bar G_0^0 $ can not be 
proportional to $\phi _1 $ nor to $\phi _2$ since the $\varphi $
tranformation  has no $\dot \epsilon $ term. So, at maximum

\begin{mathletters}
\label{56}
\begin{eqnarray}
\bar {G^0_1} &=& A \omega _1 + B \omega _2 \label{mlett:1} \\
\bar {G^0_0} &=& C \omega _1 + D \omega _2 \ \label{mlett:2} \\
 \bar { \bar {G^1}} &=& E \Phi _1 +F \phi _2 + L\omega _1 +M\omega _2 
\label{mlett:3}\\
\bar {\bar  {G^0}} &=& H \Phi _1 +I \Phi _2 +N \omega _1 +P \omega _2\,\,\, .
\label{mlett:4} \,\,\, 
\end{eqnarray}
\end{mathletters}
\noindent 
 Comparing with (\ref{55}) we find that

\begin{equation}
E=0\,\,\,\,\, , F=1 \,\,\, . \label{57}
\end{equation} 

In a similar way we obtain the $\epsilon _0$   contribution and those
  from the
metric components  $g_{\mu \nu}$
 (using (\ref{50}) and  (\ref{53})). The final result is

\begin{eqnarray}
\begin{array}{ccc}
A=2g_{01} & N=\partial _0 g_{00} & B=g_{11} \\
C=2g_{00} & R=0                  & D=g_{11}
\\
L=\partial _1 g_{00} & T=0  & M=\partial _1 g_{01}
\\
E=0 & F=1 & P=\partial _0 g_{01} \\
H= \frac{\sqrt {-g}}{g_{11}} & I=\frac{g_{01}}{g_{11}}
& S=g_{01} 
\end{array}
\label{57} \,\,\, . 
\end{eqnarray}\vskip .3cm

\noindent With the gauge generator $G$ at hand, following the ideas
of section II, we go for the calculation of the {\it generation
function}
($\bar M$)

\begin{equation}
\bar M = \int m  dy  
=  \int dy [P^i \frac{ \partial G}{\partial P^i}-G] \label{58} \,\,\, .
\end{equation}

\noindent After some algebra we obtain straightfowardly

\begin{mathletters}
 \begin{eqnarray}
m =   \frac {\sqrt {-g}}{g_{11}}
\epsilon ^0 \left[ \frac{\varphi ^2 }{2} + \frac{2}{\alpha ^2} (g_{11}
\pi ^{11})^2-\frac{\alpha g_{11}'}{2g_{11}}\varphi ' \right.\nonumber 
\\ \left. + \alpha \varphi
'' 
 + \frac{2}{\alpha } g_{11} \pi ^{11} \pi _{\varphi }\right] \nonumber 
\\ + 
(\epsilon ^1 + \frac {g_{01}}{g_{11}}\epsilon ^0 )\left[
-2 g_{11}(\pi ^{11})'-2(g_{11})'\pi ^{11}\right]
  \label{59} \,\, . 
\end{eqnarray}
\end{mathletters}

\vskip .5cm

The next step is to construct the {\it effective} hamiltonian density after
 complete gauge fixing.  We find first the reduced phase-space
   structure, using
the Dirac bracket procedure. We choose as gauge fixing constraints

\begin{equation}
\Gamma _5=\pi ^{11}-f(t) \,\,\,\,\, \Gamma _6 =\partial _1 \varphi-1 
\label{60} \,\,\, .
\end{equation}

where $f(t)$ is an arbitrary function of time.
 To obtain a more convenient form of the Dirac matrix we use the following
 linear
combinations

\begin{mathletters}
\label{61}
\begin{eqnarray}
\Lambda _1 &=&  \phi _1 +\Gamma _5 \label{mlett:1} \\
\Lambda _2 &=& \phi _1-\Gamma _5  \label{mlett:2} \,\,\, ,
\end{eqnarray}
\end{mathletters}

\noindent 
whose  Poisson brackets are

\begin{eqnarray}
\label{62}
\{\Lambda _1(x), \Lambda _1(y)\}=-2\alpha \partial _x \delta
(x-y)\,\,\, .
\end{eqnarray}

The Dirac brackets for the physical degrees of freedom can be obtained in a
two-steps procedure. First we fix the $[\pi ^{01}$-$\pi ^{00}]$ sector (
see expressions (\ref{52}) ) using the light-cone gauge fixing condition
\cite{Pol}, this is straightforward. In a second step we take the
remaining set of contraints, namely $\phi _1$, $\phi _2$, $\Lambda _1$
and $\Lambda _2$.
Finally we obtain,
 after a long
 calculation,

\begin{equation}
\{g_{11} (x), \pi ^{11} (y) \}_D= \delta (x-y)\,\,\, , \label{63}
\end{equation}

\noindent the others are zero. It is important to notice that although
the results obtained are the ``expected'' expressions (\ref{63}),
 the gravitational
 field component $g_{11}$ and its  momentum are not independent quantities,
 they are linked by the constraints relations (\ref{53}).

\vskip .5cm

 To find the {\it effective} hamiltonian in reduced phase-space we perform, as
 was explained in section II, a  canonical transformation.
 In the  gravity sector  we have

\begin{mathletters}
\label{66}
\begin{eqnarray}
\Pi ^{11} = \pi ^{11} -f(t) \\
G_{11} = g_{11} \,\,\, .
\end{eqnarray}
\end{mathletters}

\noindent The new lagrangian density reads

\begin{equation}
L' = L + \partial _{\mu } F^{\mu } \,\,\, , \label{67}
\end{equation}

\noindent where $F^{\mu}$ is the generator of the canonical transformation. The
correct equations of motion are obtained when

\begin{equation}
\label{68}
F^0=\Pi ^{11} g_{11} \,\,\,\,\,\, F^1= \alpha \left(
1-\frac{1+g_{11}}{2g_{11}}
\right) \partial _1 g_{11}\Pi ^{11} \,\,\, .
\end{equation}

\noindent Finally, the {\it effective} hamiltonian density is easily computed

\begin{equation}
H_{eff}=g_{11}+ \alpha \left( 1-\frac{1+g_{11}}{2g_{11}} \right)
\Pi ^{11}\partial
_1 g_{11} \,\,\, , \label{69}
\end{equation}

\noindent  and this density rules the dynamics of the gravitational field in the
  reduced
 phase-space. In fact, this result is in  accord  with the time
 derivative of  $g_{11}$ (obtained from the definition of momenta and the
gauge fixing conditions (\ref{60}) )
 
\begin{equation}
 \dot g_{11}(x) = \left[ 4\frac {g_{11}}{\alpha ^2} + 2\left(
1- \frac{g_{11}+1}{
2g_{11}}\right) \partial _1 g_{11}\right](x) \,\,\,\,\, . 
  \label{70}
 \end{equation}

\section {Conclusions}

The methods developed by Henneaux, Teitelboim and Vergara, and independently
by Gitman and Tiutyn, offer a solution to the  zero hamiltonian problem
(ZH problem) in the 2-D induced gravity case. They permit to obtain
an {\it effective } hamiltonian, which rules  the evolution of physical
 degrees of freedom after complete gauge fixing. The key point is that open
  boundary
 conditions are necessary; the new hamiltonian  arises naturally after a 
time-dependent canonical transformation is performed in reduced phase-space.

{\bf Acknowledgements} 

The authors would like to Thank  Capes (Brazil) and the Physics
Department of UFES (Vit\'oria-Brazil) for finnancial support.

\end{document}